\documentstyle[epsf,twocolumn,seceq]{orb2001}

\title
{
Antiferroquadrupolar Order in the Magnetic Semiconductor TmTe
}

\def\runtitle
{
Antiferroquadrupolar Order in TmTe
}

\author
{
Jean-Michel {\sc Mignot},
Arsen {\sc Gukasov},
Changping {\sc Yang},
Peter {\sc Link}\footnote{Present address: Inst. f\"{u}r 
Physikalische Chemie, Universit\"{a}t G\"{o}ttingen, Tammannstr. 6, 
G\"{o}ttingen, Germany.},
Takeshi {\sc Matsumura}$^{1}$ and
Takashi {\sc Suzuki}$^{2}$
}

\def\runauthor
{
Jean-Michel {\sc Mignot} {\it et al.}
}

\inst
{
Laboratoire L\'{e}on Brillouin, CEA-CNRS, CEA/Saclay, F-91191 Gif sur 
Yvette, France\\
$^{1}$Department of Physics, Tohoku University, Sendai 980-77, Japan\\
$^{2}$National Research Institute for Metals, Tsukuba, Ibaraki 
305-0047, Japan

}

\abst
{
The physical properties of the antiferroquadrupolar state occurring in 
TmTe below $T_{\mathrm{Q}}=1.8$~K have been studied using neutron 
diffraction in applied magnetic fields. A field-induced antiferromagnetic component 
(${\mib k}=(\frac{1}{2},\frac{1}{2},\frac{1}{2})$) is observed and, 
from its magnitude and direction for different orientations of $H$, 
an $O_{2}^{2}$ quadrupole order parameter is inferred. Measurements 
below $T_{\mathrm{N}}\approx 0.5$~K reveal that the magnetic 
structure is canted, in agreement with theoretical predictions for in-plane 
antiferromagnetism. Complex domain repopulation effects occur when the 
field is increased in the ordered phases, with discontinuities in the 
superstructure peak intensities above 4~T.}

\kword
{
quadrupole order, antiferroquadrupolar, antiferromagnetic, neutron 
diffraction, TmTe
}

\begin{document}
\sloppy
\maketitle


\section{Introduction}
The number of experimental studies devoted to orbital 
phenomena in magnetic materials has grown rapidly during the last five 
years.  Long-range order involving orbital degrees of freedom has 
now been reported for a broad range of materials, from 
transition-metal oxides (V$_{2}$O$_{3}$, LaMnO$_{3}$, etc.) to 
rare-earth or actinide compounds. In contrast to 3\textit{d} elements,
lanthanide ions are characterized by a strong coupling between the total spin 
and orbital momenta, which rules out pure orbital order. Charge order, an 
important issue in manganites, is also not relevant for ``normal"rare-earth 
compounds because the occupancy of the 4\textit{f} shell has a fixed integral 
value. On the other hand, it has long been recognized 
\cite{Blume_362, Sivardiere_363}
that interactions between multipole moments of the 
\mbox{\textit{f}-electron} wavefunctions --- especially the quadrupole (QP) 
moments representing the asphericity of the charge distribution --- can 
play a significant role at low temperature, and even lead to a phase 
transition into a long-range ordered state.\cite{Morin_231} 
In analogy with magnetic systems, 
this state is termed ``ferroquadrupolar" when the  order 
parameter is uniform, ``antiferroquadrupolar" (AFQ) when its 
periodicity differs from that of the lattice.\cite{foot:1}

A few members of the second class have been known for as much as 20 
years (CeB$_{6}$,\cite{Peysson_758} PrPb$_{3}$\cite{Bucher_760}), 
but more examples have been discovered recently 
(DyB$_{2}$C$_{2}$\cite{Yamauchi_488}) as a result of an intensive 
experimental effort using a broad spectrum of techniques. 
The fingerprint of the AFQ transition may be elusive in the
magnetic susceptibility,\cite{Matsumura_251} but clear anomalies are 
observed in the specific heat 
\cite{Peysson_758, Matsumura_251} 
or the elastic constants.\cite{Matsumura_251, Goto_765} In most cases, the 
complexity arising from the tensor character of the QP moment operator 
can be disentangled only through single-crystal diffraction measurements.

In this context, \mbox{x-ray} scattering has the unique 
advantage that it can probe the asphericity of the 4\textit{f} charge 
distribution without the need to apply a magnetic field. AFQ 
order was detected recently by this technique in several compounds
(NdMg,\cite{Amara_766} 
CeB$_{6}$,\cite{Nakao_771, Yakhou_770} 
UPd$_{3}$,\cite{McMorrow_769} 
DyB$_{2}$C$_{2}$\cite{Hirota_768, Tanaka_767}).
However, such experiments are not straightforward because the 
measured intensities are very weak in the case of conventional
Thomson scattering, and those obtained in  resonant experiments 
are difficult to interpret quantitatively. Neutrons, which have no direct interaction 
with QP moments, can nonetheless disclose the ``hidden" QP order 
through its effect on the magnetic response to an applied magnetic field. 
This was first demonstrated in the case of 
CeB$_{6}$ by Effantin {\it et al.},\cite{Effantin_232} 
who observed field-induced antiferromagnetic (AFM) peaks below the transition 
temperature $T_{\mathrm{Q}}$, reflecting the existence of distinct easy axes on 
the two AFQ sublattices. That strategy also proved quite successful 
with TmTe, as reported below. Furthermore, neutron provide an ideal tool 
to study the magnetic order which sets in at lower temperature and the interplay 
between the QP and dipole order parameters.

TmTe is a cubic, NaCl structure, magnetic semiconductor with an energy 
gap of about 0.35~eV.\cite{Wachter_216} Tm ions are divalent 
(4\textit{f}$^{13}$), and the ground-state multiplet $^{2}F_{7/2}$ is 
split by the crystal field into one quartet ($\Gamma_{8}$) and two 
Kramers doublets ($\Gamma_{7}$) and ($\Gamma_{6}$). Neutron 
scattering results\cite{Clementyev_735} indicate that the overall splitting 
is quite small (about 1~meV) and that the ground state is most likely 
$\Gamma_{8}$, as proposed previously from elastic constant and 
thermal expansion measurements.\cite{Ott_220, Ott_772}
AFQ order, with a transition temperature of  $T_{\mathrm{Q}}=1.8$~K 
was discovered by Matsumura {\it et al.}\cite{Matsumura_217, Matsumura_251}
using specific heat and ultrasonic experiments. The study of QP 
interactions in this system is of particular interest because, 
unlike intermetallic compounds such as CeB$_{6}$ or PrPb$_{3}$, 
TmTe has a very low carrier concentration at temperatures of the order of 
$T_{\mathrm{Q}}$, and the role of interactions mediated by conduction 
electrons\cite{Levy_364} can be considered negligible. On the 
other hand, no sizeable lattice distortion indicative of magnetoelastic 
couplings could be detected so far. AFM order was reported to occur 
with a transition temperature $T_{\mathrm{N}}$ comprised between 
0.23~K (magnetic susceptibility
\cite{Matsumura_251, Bucher_759, Sakakibara_790} 
and 0.43~K (neutron diffraction).\cite{Lassailly_221}

In the following, we report a detailed study of the AFQ and AFM 
phases of TmTe by neutron diffraction. Section \ref{subsec:afqo} 
presents the observation of the field-induced AFM superstructure  
below $T_{\mathrm{Q}}$, and the analysis of the data for different 
field directions, leading to the identification of the order parameter.
These results have been described in more detail in previous publications.
\cite{Link_252, Link_358, Link_383, Mignot_424}
Next we set out to discuss the magnetic transition and the dipole-ordered 
phase occurring below $T_{\mathrm{N}}$. Recent measurements carried 
out down to 100~mK are reported in Section \ref{subsec:dmo}. 
They reveal the existence of a canting of the magnetic structure which can 
be regarded as the consequence of QP order. The application of large magnetic 
fields at very low temperature produce domain reorientation phenomena with 
dramatic irreversibilities. The final section outlines possible directions 
for future studies on this system.


\section{Experiments}
To observe QP order in TmTe and determine its phase diagram, it is 
necessary to apply large magnetic fields and to cool the 
sample down to $T<T_{\mathrm{Q}}=1.8$~K. Even lower temperatures 
are required for studying the transition into the AFM phase. 
These conditions were achieved by using two different split-coil, vertical field 
cryomagnets with maximum fields of 7.5 and 12~T, respectively. 
The former could be equipped with a \mbox{$^{3}$He-$^{4}$He} 
dilution insert yielding a minimum mixing-chamber temperature of about 
100~mK. Thermal coupling of the sample to the cold plate carrying the 
temperature sensors was ensured by a gold wire, which proved to be 
effective down to about 150~mK.

Neutron diffraction experiments were carried out  on the two-axis 
lifting-counter diffractometers 6T2 (thermal beam) and 5C1 (hot 
source) at the Orph\'{e}e reactor in Saclay. Measurements were made 
at neutron wavelengths of 0.90 or 2.35 \AA (6T2), and 0.84 \AA (5C1) 
with a primary beam collimation of 50'.  Second-order contaminations 
were suppressed by means of erbium or pyrolytic graphite filters. For the
polarized-neutron experiments, 5C1 was operated with a Heusler-alloy 
monochromator, providing a polarization $P_{0}$ = 0.91 at 0.84 \AA.  

Measurements were taken on a large TmTe single crystal 
(lattice constant $a_{0}$ = 6.354 \AA), prepared by induction-melting of 
high-purity constituents inside a vacuum-sealed
tungsten crucible.\cite{Matsumura_251} From its lattice 
constant, as well as from polarized-neutron results reported previously,
\cite{Link_358} this sample was concluded to be close to 
stoichiometric, with an estimated Tm deficiency of $0.04\pm 0.01$. 
Different orientations of the crystal with respect to the magnetic field 
(${\mib H} \parallel$ [110], [001] and [111]) were studied in separate experiments.


\section{Results}\label{sec:res}

\subsection{Antiferroquadrupolar order}\label{subsec:afqo}
\begin{figure}[t]
     \epsfxsize=7.5cm
      \epsfigure{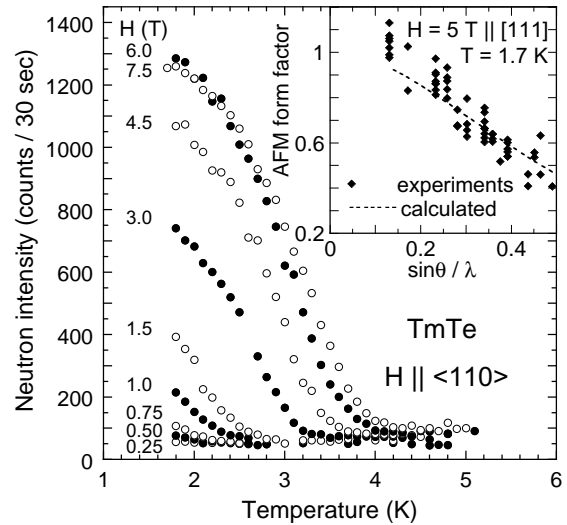}
  \caption{Intensity of the field-induced magnetic Bragg peak 
  \mbox{${\frac{1}{2}\frac{1}{2}\frac{1}{2}}$}
  as a function of temperature for different magnetic fields applied 
  along [110]. Inset: magnetic form factor associated with the induced 
  AFM reflections measured at $T=1.7$~K in a field of 5~T along [111]; 
  the dashed line represents theTm$^{2+}$ form factor calculated in the 
  dipolar approximation.}
  \label{fig:1}
\end{figure}

When TmTe is cooled in an applied magnetic field, 
superstructure peaks corresponding to the wave vector 
${\mib k}=(\frac{1}{2},\frac{1}{2},\frac{1}{2})$ (\mbox{type-II} AFM) appear 
below the critical temperature $T_{\mathrm{Q}}$. Intensities measured 
with ${\mib H} \parallel [110]$ are shown in Fig.~\ref{fig:1} for different field 
values. The reflections vanish for $H=0$, which indicates that the 
transition occurring in zero field is not associated with magnetic 
dipole order. On the other hand, the ${\mib Q}$ dependence of the 
field-induced intensities approximately follows that calculated for the 
$\mathrm{Tm}^{2+}$ magnetic form factor (inset in Fig.~\ref{fig:1}). 
Therefore the observed superstructure results from magnetic 
scattering and cannot be ascribed to a distortion of the lattice, 
which would at any rate require unphysically large atomic 
displacements to account for the 
experimental peak intensities. The observation of a staggered 
magnetic response in a uniform applied field is explained naturally by 
assuming that the transition at $T_{\mathrm{Q}}$ corresponds to the ordering 
of 4\textit{f} QP moments in a long-range structure with the wave 
vector $(\frac{1}{2},\frac{1}{2},\frac{1}{2})$, which splits the Tm 
sites into two sublattices with different local anisotropy axes. 
Similar results have been obtained for ${\mib H} \parallel [111]$.  For 
${\mib H}\parallel [001]$ extra peaks are also observed but their intensities
are much weaker.  Phase diagrams drawn from the neutron data for the
three field directions are in general agreement with those derived
previously from the specific heat.  In particular, the critical
temperature is strongly enhanced by the external field for all 
symmetry directions, and the QP transition line closes up above 5~T for 
${\mib H} \parallel [001]$.

\begin{figure}[t]
     \epsfxsize=7.5cm
    \epsfigure{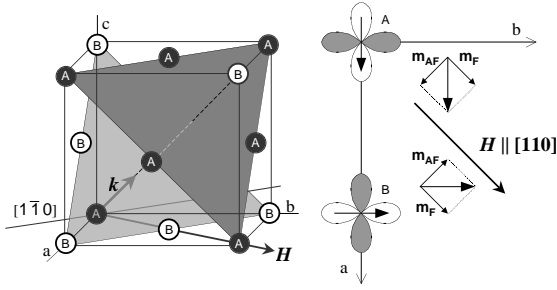}
  \caption{Schematic representation of the AFQ structure
  showing the A and B sublattices associated with the wave vector 
  ${\mib k_{1}}$ (left), and the field-induced uniform (${\mib {m}_{\mathbf{F}}}$) and 
  staggered (${\mib {m}_{\mathbf{AF}}}$)  components 
  for ${\mib H} \parallel [110]$ (right).}
  \label{fig:2}
\end{figure}

To obtain more detailed information on the AFQ order we have
refined the induced AFM structure for ${\mib H} \parallel [110]$ and 
${\mib H} \parallel [111]$. For both orientations, the AFM Fourier 
component ${\mib {m_{k}}}$ was found to be directed along one of the  
three two-fold cubic axes perpendicular to the wave vector ${\mib k}$.
This gives rise to three possible magnetic domains, denoted \mbox``{$S$-domains}",
whose populations can be derived from the refinement.
From the field dependence of the \mbox{$k$-domain} populations (see below) 
the structure is expected to be \mbox{single-${\mib k}$}, 
and the the AFM moment on the Tm sublattices is thus 
equal to $\pm{\mib {m_{k}}}$. The resulting structure for ${\mib H} \parallel 
[110]$ is represented in Fig.~\ref{fig:2} for the wave vector 
$(\frac{1}{2},\frac{1}{2},\frac{1}{2})$. The AFM component 
(${\mib {m_{k}}}=1.6~\mu_{\mathrm{B}}$ for $H=5~\mathrm{T}$) is drawn 
perpendicular to both ${\mib H}$ and ${\mib k}$ because this 
corresponds to the \mbox{$S$-domain} found to prevail in high fields. 
Also shown is the uniform magnetic component along ${\mib H}$ 
(${\mib m_{0}}=2.60(3)~\mu_{\mathrm{B}}$ for $H=4~\mathrm{T}$ from the 
polarized-neutron results). For ${\mib H} \parallel [001]$, 
a similar refinement was not possible because the experimental 
intensities were not strong enough.

Using the symmetry relations between the AFQ order parameter and 
the field-induced dipole moments derived previously for 
CeB$_{6}$,\cite{Shiina_234} it can be 
inferred from the above results that the primary QP order parameter is most 
likely $O_{2}^{2}$.\cite{Link_252} A detailed study of multipole interactions 
in TmTe has been carried out by Shiina {\it et al.}\cite{Shiina_779} 
using a classical mean-field model, in which the Tm sites are 
considered to form four uncoupled simple-cubic (\textit{sc}) sublattices.  Their
results indeed confirm that, if the crystal field is chosen with an
easy axis along [001], the experimental data can be explained
consistently in terms of a $\Gamma_{3}$ (tetragonal) QP order parameter. 
The situation for ${\mib H} \parallel [110]$ is depicted schematically in 
Fig.~\ref{fig:2}.

\begin{figure}[t]
     \epsfxsize=8.5cm
      \epsfigure{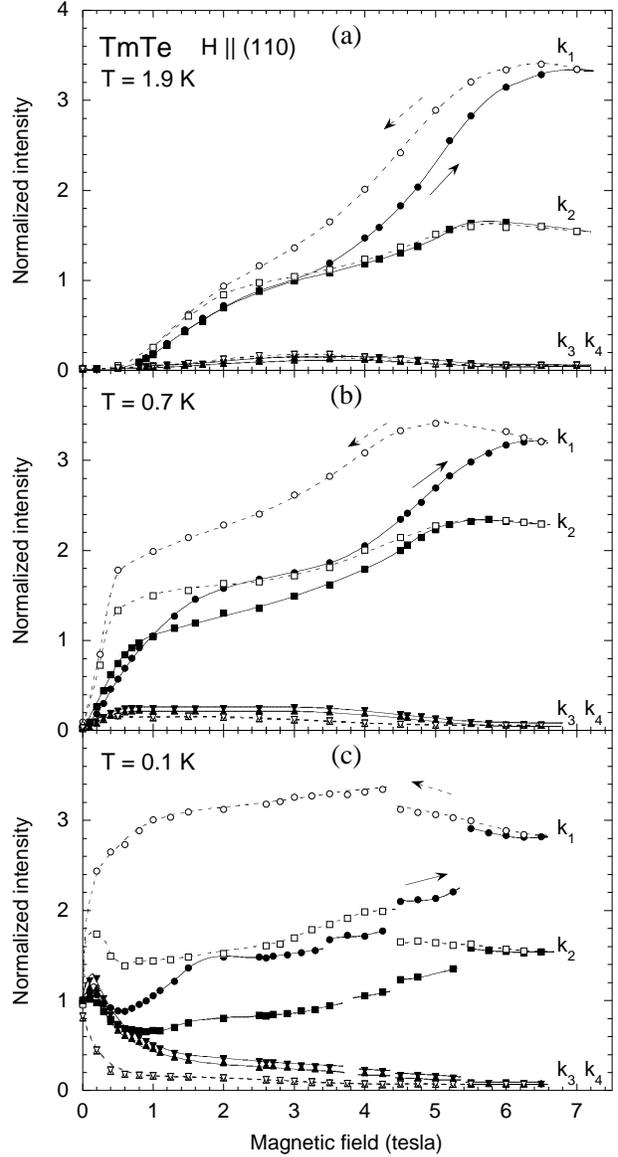}
  \caption{Intensities of the magnetic peaks 
  ${\pm\frac{1}{2}\pm\frac{1}{2}\pm\frac{1}{2}}$  associated with 
  the four \mbox{$k$-domains} ${\mib k_{1}}\ldots{\mib k_{4}}$
  as a function of the applied field $H\parallel [110]$, measured in 
  the AFQ (a, b) and AFM (c) phases. Closed (open) symbols denote 
  measurements taken on increasing (decreasing) $H$. 
  Data for each ${\mib k}$ vector have been normalized to the zero-field 
  value at $T=0.1$~K (see Section \ref{subsec:dmo}).}
  \label{fig:3}
\end{figure}

The variations of the AFM peak intensities with the external field reveal 
interesting domain repopulation effects. In low fields,
the four different \mbox{$k$-domains}, associated with the wave vectors 
${\mib k_{1}}=(\frac{1}{2},\frac{1}{2},\frac{1}{2})$, 
${\mib k_{2}}=(\frac{1}{2},\frac{1}{2},-\frac{1}{2})$, 
${\mib k_{3}}=(\frac{1}{2},-\frac{1}{2},\frac{1}{2})$ and 
${\mib k_{4}}=(-\frac{1}{2},\frac{1}{2},\frac{1}{2})$ 
are found to exist. If $H$ is increased along [111], one single 
domain, corresponding to ${\mib k_{1}} \parallel {\mib H}$, grows at the 
expense of the other three. This behavior implies that the structure 
is \mbox{single-${\mib k}$}. A different situation 
exists for  ${\mib H} \parallel [110]$ as \textit{two} domains, 
${\mib k_{1}}=(\frac{1}{2},\frac{1}{2},\frac{1}{2})$ and 
${\mib k_{2}}=(\frac{1}{2},\frac{1}{2},-\frac{1}{2})$, 
are favored in high fields (Fig.~\ref{fig:3}).  As in the previous
case, the dominant domains turn out to be those whose ${\mib k}$
vector makes the smallest angle (35 degrees) with the external
field.  The above model,\cite{Shiina_779} completed by introducing an 
additional electrostatic \mbox{QP-QP} coupling between the \textit{sc}
sublattices,\cite{Shiina_788} correctly predicts  the 
\mbox{single-${\mib k}$} structure, as well as the domain repopulation 
for ${\mib H} \parallel [111]$, but not that for ${\mib H} \parallel [110]$.  
A more realistic treatment of these interactions appears to be necessary.

The field dependences (${\mib H} \parallel [110]$) of the induced intensities  of 
\mbox{$\pm\frac{1}{2}\pm\frac{1}{2}\pm\frac{1}{2}$} reflections 
corresponding to the four \mbox{$k$-domains} are plotted in 
Figs.~\ref{fig:3}a and \ref{fig:3}b
for two temperatures within the AFQ phase.
At $T=1.9$~K the zero-field state is paramagnetic and the AFM signal 
therefore appears only above a threshold at about 0.5~T. At $T=0.7$~K, 
on the other hand, the intensity increases steeply from $H=0$.
In both cases the variation takes place in two steps, and the 
intensities from the depleted domains ${\mib k_{3}}$ and ${\mib k_{4}}$
start to decrease only above 3.5~T, in connection with a pronounced 
upturn in those from the dominant domains. As a result, a 
sizeable AFM component is induced even in the minority domains before 
the \mbox{$k$-domain} repopulation is completed. Another point to be 
noted is that the intensities associated with ${\mib k_{1}}$ and 
${\mib k_{2}}$ differ significantly in large fields, especially at 
$T=1.9$~K, presumably because a small deviation from the nominal
sample orientation with respect to the field makes these domains not 
strictly equivalent. In contrast, their variations remain very similar up to 3~T.
A \mbox{double-${\mib k}$} to \mbox{single-${\mib k}$} transition 
could produce this type of behavior, but it seems incompatible with the 
domain repopulation observed for ${\mib H} \parallel [111]$. Another 
possibility is that the \mbox{$k$-domains} essentially retain 
their initial populations up to 3~T, and that the difference between 
the two types of domains reflects their intrinsic staggered response 
in the applied field. Detailed data collections and structure 
refinements at several values of $H$ should be conducted to clarify this point.
As could be expected, large irreversibilities at increasing and 
decreasing field are observed in connection with the changes in 
domain populations. This effect becomes more pronounced at
lower temperature.

\subsection{Dipole moment order}\label{subsec:dmo}
Various experimental results indicate that a second 
phase transition corresponding to the ordering of the Tm dipole 
moments, takes place at $T_{\mathrm{N}} \ll T_{\mathrm{Q}}$.
\cite{Matsumura_251, Bucher_759, Sakakibara_790, Lassailly_221} 
The magnetic structure was identified as  \mbox{type-II} AFM from the 
observation of neutron Bragg peaks associated with 
the wave vector ${\mib k}=(\frac{1}{2},\frac{1}{2},\frac{1}{2})$.
\cite{Lassailly_221}
At the time that result was reported, the existence of AFQ order was 
not suspected and it was therefore natural to assume the magnetic 
structure to be conventional AFII. However, the mean-field 
theory\cite{Shiina_779} suggests that the AFQ order should have 
sizeable effects on the magnetic phase transition: it is expected to favor 
FM correlations, thereby causing a substantial reduction of $T_{\mathrm{N}}$ 
and, in the case of  \mbox{``in-plane "} AFM order 
(ordered dipole moments lying within the plane of the $O_{2}^{2}$ quadrupoles)
 which seems appropriate for TmTe, to produce a canting of the AFM 
 structure. This canting consists in the rotation of the dipole 
 moments on each sublattice by $+\phi$ and $-\phi$ respectively with respect 
 to the local anisotropy axis, and should be observable experimentally.
 
 \begin{figure}[t]
     \epsfxsize=6.5cm
      \epsfigure{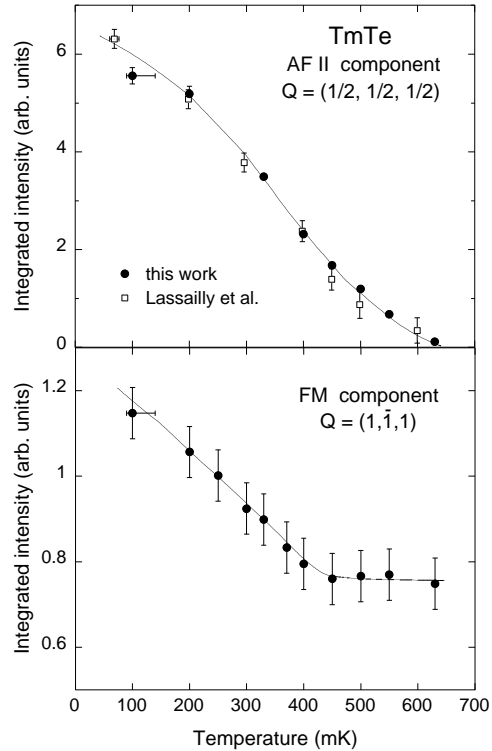}
  \caption{
  Integrated intensities of the $\frac{1}{2}\frac{1}{2}\frac{1}{2}$ 
  (AFM) and $1\bar{1}1$ (FM) peaks measured in zero field as a 
  function of temperature. The data of ref.~25 are plotted in the 
  upper frame for comparison.}
  \label{fig:4}
\end{figure}

 Our measurements for $H=0$ confirm the existence of AFII magnetic Bragg 
 peaks. In the upper frame of Fig.~\ref{fig:4}, the integrated intensity of the 
\mbox{ $\frac{1}{2}\frac{1}{2}\frac{1}{2}$} reflection is plotted 
 as a function of $T$, together with the previous results of Lassailly 
 {\it et al.}\cite{Lassailly_221} Despite the use of crystals from different 
 origins, the agreement between the two sets of data is excellent. A tendency to 
 saturation is observed in our measurements below 0.15~K, probably due to 
 insufficient thermalization of the sample exposed to the neutron beam.
 One notes that the magnetic intensity persists far above 
 the N\'{e}el temperature $T_{\mathrm{N}}=0.45$~K
 reported in ref.~25. However, the sharp maximum in the $T$ dependence of the 
 Bragg peak \textit{width} on which the latter estimate was based could not be reproduced 
 in the present experiments. Our data merely indicate a moderate monotonic
 broadening of the rocking curves accompanying the intensity drop as the 
 system is heated up to 0.6~K. The determination of $T_{\mathrm{N}}$ from 
 the neutron intensities thus remains elusive. We recall that similar difficulties 
 were also encountered with other techniques, as the ac susceptibility
 \cite{Matsumura_251, Bucher_759, Sakakibara_790}  
 shows a divergence at 0.22~K but no anomaly around 0.5~K, and the 
 specific-heat peak corresponding to the AFM transition exhibits a complex 
 structure with a peak at 0.34~K followed by a shoulder 
 near 0.5~K.\cite{Matsumura_251} These problems may reflect the 
 existence of anomalously strong fluctuations, which have been predicted to 
 occur because of competing interactions between different multipole moments. 

The lower frame in Fig.~\ref{fig:4} represents the integrated 
intensity of the weak $1\bar{1}1$  nuclear peak as a function of temperature. 
The increase observed below 0.45~K clearly demonstrates that a 
superimposed ferromagnetic component develops below $T_{\mathrm{N}}$ 
in addition to the AFM superstructure, with a magnitude of about 
0.4~$\mu_{\mathrm {B}}$, as compared to 2.3~$\mu_{\mathrm{B}}$ 
for the AFM component. The observation of this canting component 
lends additional support to the model of ref.~31. Within the limited 
accuracy of the measurement, no significant magnetic signal seems to 
remain above 0.45~K, in contrast to the AFM component.

\begin{figure}[t]
     \epsfxsize=6.5cm
      \epsfigure{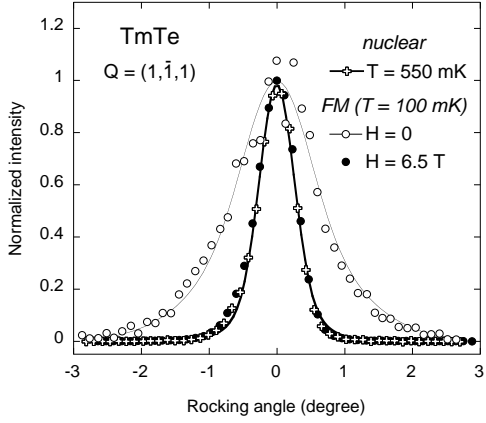}
  \caption{
  Normalized rocking curves through the $1\bar{1}1$ reflection. 
  Crosses and thick lines: nuclear contribution with Gaussian fit; 
  open circles and thin line: zero-field FM component with Gaussian fit; 
  closed circles: high-field FM component.}
  \label{fig:5}
\end{figure}

Indications as to the dimensions of the magnetic domains can be derived from 
an analysis of the width of the $1\bar{1}1$ reflection. At $T=0.55$~K in zero 
field, the intensity is purely nuclear and the width of the rocking 
curve thus reflects the experimental resolution and the mosaic of the crystal. 
For $T=0.1$~K, the magnetic intensities were obtained by subtracting out 
this nuclear signal. The results for $H=0$ and 6.5~T are plotted in
Fig.~\ref{fig:5}, together with the nuclear peak. For the sake of comparison, 
all curves have been normalized because their intensities differ by 
orders of magnitude (nucl:FM[$H=0$]:FM[6.5~T] $\approx$ 5:1:150). 
The graph clearly shows that the magnetic peak in zero field is substantially 
broader then the resolution. This effect can be ascribed to a finite 
correlation length $\xi$ of the FM component, estimated to be of the order of 
50~\AA, which probably reflects the existence of magnetic domains with 
different directions of canting.
In contrast, the FM component measured at 6.5~T can be superimposed 
almost perfectly on the nuclear peak, implying that $\xi$ becomes 
infinite (to the precision of the measurement) when the field becomes 
large enough to reorient the domains. This effect takes place at relatively low fields, 
as a narrow peak is already recovered at $H=2$~T.

The field dependences (${\mib H} \parallel [110]$) of the AFM peak intensities for 
$T\approx 0.1~{\mathrm{K}} < T_{\mathrm{N}}$ is displayed in Fig.~\ref{fig:3}c. 
In contrast to the data for $T > T_{\mathrm{N}}$ plotted in frames a and 
b,  finite intensities exist already in zero field because the wave vectors 
of the AFM and AFQ ordered states are the same. The measured values  
are different for ${\mib k_{1}}$ and ${\mib k_{2}}$ on the one hand, 
${\mib k_{3}}$ and ${\mib k_{4}}$ on the other hand, owing to 
different scattering geometries (Lorentz factor and resolution). 
However, it was checked by comparing integrated intensities for 
different sets of equivalent reflections that the zero-field 
populations of the four \mbox{$k$-domains} do not differ 
significantly. The curves in Fig.~\ref{fig:3}c were thus normalized 
to the same value for $H=0$, and the same scaling factor was also 
applied to the corresponding data plotted in the other two frames. 

 \begin{figure}[t]
     \epsfxsize=6.5cm
      \epsfigure{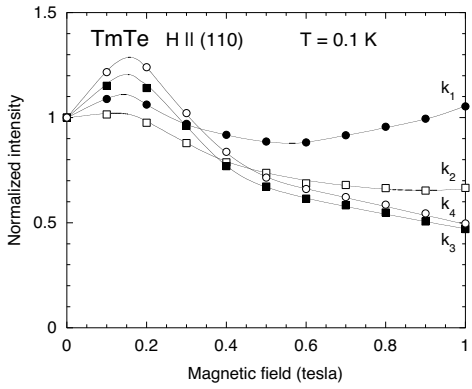}
  \caption{Intensities of the magnetic peaks 
  ${\pm\frac{1}{2}\pm\frac{1}{2}\pm\frac{1}{2}}$  associated with 
  the four \mbox{$k$-domains} ${\mib k_{1}}\ldots{\mib k_{4}}$
  as a function of the applied field $H\parallel [110]$, measured in the AFM 
  phase (expanded plot of the low-field part of Fig.~\ref{fig:3}c).}
  \label{fig:6}
\end{figure}

One can roughly distinguish three different regimes as a function of 
the applied field. Below 1~T (Fig.~\ref{fig:6}), a general decrease in the AFM 
intensity takes place, with initially a weak maximum near 0.15~T for 
${\mib k_{3}}$ and ${\mib k_{4}}$. This region is dominated by dipole moment 
order and may involve some reorientation among \mbox{$S$-domains}. Such 
a view is consistent with the specific-heat results, in which a magnetic-ordering 
anomaly could be traced up to $H \approx 1$~T. At higher fields, the 
intensities for the ${\mib k_{1}}$ and ${\mib k_{2}}$ domains go 
through a minimum and start to increase. The  situation resembles that 
observed at $T=0.7$~K, \textit{i.e.} for pure AFQ order. However, for 
fields in excess of 3~T, one observes a succession of jumps, with a 
positive sign for ${\mib k_{1}}$ and ${\mib k_{2}}$, a negative sign and a 
smaller magnitude for ${\mib k_{3}}$ and ${\mib k_{4}}$. Similar 
discontinuities, known as ``Barkhausen steps "are commonly observed in the 
magnetization curves of ferromagnets. 
They originate from random irreversible domain-wall jumps as $H$ 
is increased. In Section \ref {subsec:afqo}, it was indeed suggested that most 
of the repopulation of AFQ domains takes place in the range $H > 3$~T. 
The observation of such a behavior in an AFM phase is quite interesting 
and calls for further investigations. Surprizingly, no anomalies have been 
detected in the bulk magnetization for this field direction.
\cite{Sakakibara_790} How domain reorientation 
responsible for sizeable jumps in the AFM field-induced component can 
have no fingerprint in the uniform component thus remains to be explained.


\section{Conclusions}
The results obtained in the present work provide a comprehensive 
picture of ordering phenomena in TmTe, involving both dipole and QP 
moments. Qualitative agreement is found with the main predictions of a 
mean-field  model based on a classical treatment of crystal-field, 
quadrupolar and exchange interactions. In particular, the canting of 
the magnetic moments anticipated for in-plane AFM order is observed 
experimentally below $T_{\mathrm{N}}$. On the other hand, a number of 
questions remain open. Probably the most important one concerns the nature 
of QP interactions in TmTe. As noted at the outset, the 
conduction-electron-mediated, \mbox{RKKY-type}, 
mechanism\cite{Levy_364} proposed 
for intermetallic compounds is not applicable here. Alternative 
possibilities are a coupling through the lattice, as in Jahn-Teller 
insulators,\cite{Gehring_791} or a superexchangelike interaction 
between Tm next nearest neighbors 
separated by one Te atom as proposed in ref.~31. Experimentally, it is 
not simple  to distinguish between those mechanisms: in particular, it 
has been argued that AFQ order itself should result in sizeable 
lattice distortions below $T_{\mathrm{Q}}$.\cite{Nikolaev_793} High-resolution \mbox{x-ray} 
diffraction experiments should be performed to clarify this point. 
Resonant experiments could also allow the AFQ order to be probed directly 
in zero field. This requires, however, to reach temperatures of less than 1.8~K, 
and to obtain crystals with very small mosaic spread.

Our present interpretation of the AFQ phase as due to an $O_{2}^{2}$ 
QP order parameter seems consistent with the bulk of experimental 
observations, but some points require further examination. In 
particular, the results for ${\mib H} \parallel [001]$ are not well 
understood, mainly because of the weakness of the signals. Even 
so, the fact that a finite induced AFM component exists seems in 
contradiction with the symmetry arguments of Shiina {\it et al.}
\cite{Shiina_234}
which predict that component to vanish. This may indicate 
that the latter analysis, initially performed for  CeB$_{6}$ ({\it 
sc}) is not readily applicable to {\it fcc} TmTe because the small groups of 
the ${\mib k}$ vector $(\frac{1}{2},\frac{1}{2},\frac{1}{2})$ for the 
two types of Bravais lattices are different.\cite{Sakai_000} It was also predicted 
that for this field direction, the external field should stabilize the component 
$O_{2}^{0}$ with respect to $O_{2}^{2}$, possibly causing a 
transition between two AFQ phases.\cite{Shiina_779} 
Experimental evidence for such a transition is still lacking but it has 
been suggested that it might be responsible for the steplike jump 
observed in the bulk magnetization for ${\mib H} \parallel [001]$ 
at very low temperature.\cite{Sakakibara_790}
As to the properties of the AFM state, the results reported in this 
paper reveal quite interesting behaviors, especially in strong applied 
magnetic fields. Additional experiments are required to work out a 
reliable refinement of the magnetic structure in the presence of 
complex domain reorientation effects.

\section*{Acknowledgements}
We thank Th.\ Beaufils, Ph.\ Boutrouille and \mbox{P.\ Fouilloux }
for help with the experiments, \mbox{R. Shiina }for numerous comments 
and suggestions, and A. V. Nikolaev, H. Shiba and O. Sakai for useful 
discussions.



\begin{thebibliography}{99}

\bibitem{Blume_362} M. Blume and Y. Y. Hsieh: J. Appl. Phys. {\bf 40} (1969) 1249.
\bibitem{Sivardiere_363} J. Sivardi\`{e}re and M. Blume: Phys. Rev. B {\bf 5} (1972) 1126.
\bibitem{Morin_231} P. Morin and D. Schmitt: {\it Ferromagnetic Materials}, ed. K. H. J. Buschow and E. P. Wohlfarth (North-Holland 1990) p. 1.
\bibitem{foot:1} Strictly speaking, one should distinguish between different types of non-uniform QP structure (commensurate / incommensurate, single- / \mbox{multi-${\mib k}$}, collinear / noncollinear). However, experimental evidence is so far lacking for the occurrence of such situations in real systems.
\bibitem{Peysson_758} Y. Peysson, C. Ayache, J. Rossat-Mignod, S. Kunii and T. Kasuya: J. Phys. (France) {\bf 47} (1986) 113.
\bibitem{Bucher_760} E. Bucher, K. Andres, A. C. Gossard and J. P. Maita: {\it Proceedings of the 13th Conference on Low Temperature Physics, LT 13}, ed. K. D. Timmerhaus, W. J. O'Sullivan and E. F. Hammel (Plenum Publishing Corp., New York 1974) p. 322.
\bibitem{Yamauchi_488} H. Yamauchi, H. Onodera, K. Ohoyama, T. Onimaru, M. Kosaka, M. Ohashi and Y. Yamaguchi: J. Phys. Soc. Jpn. {\bf 68} (1999) 2057.
\bibitem{Matsumura_251} T. Matsumura, S. Nakamura, T. Goto, H. Amitsuka, K. Matsuhira, T. Sakakibara and T. Suzuki: J. Phys. Soc. Jpn. {\bf 67} (1998) 612.
\bibitem{Goto_765} T. Goto, A. Tamaki, S. Kunii, T. Nakajima, T. Fujimura, T. Kasuya, T. Komatsubara and S. B. Woods: J. Magn. 
Magn. Mater. {\bf 31} (1983) 419.
\bibitem{Amara_766} M. Amara, R. M. Gal\'{e}ra, P. Morin and J. F. B\'{e}rar: J. Phys.: Condens. Matter {\bf 10} (1998) L743.
\vspace{20cm}
\bibitem{Nakao_771} H. Nakao, K. I. Magishi, Y. Wakabayashi, Y. Murakami, K. Koyama, K. Hirota, Y. Endoh and S. Kunii: J. Phys. Soc. Jpn. {\bf 70} (2001) 1857.
\bibitem{Yakhou_770} F. Yakhou, V. Plakhty, H. Suzuki, S. Gavrilov, P. Burlet, L. Paolasini, C. Vettier and S. Kunii: Phys. Lett. A {\bf 285} (2001) 191.
\bibitem{McMorrow_769} D. F. McMorrow, K. A. McEwen, U. Steigenberger, H. M. R¿nnow and F. Yakhou: Phys. Rev. Lett. {\bf 87} (2001) 057201/1.
\bibitem{Hirota_768} K. Hirota, N. Oumi, T. Matsumura, H. Nakao, Y. Wakabayashi, Y. Murakami and Y. Endoh: Phys. Rev. Lett. {\bf 84} (2000) 2706.
\bibitem{Tanaka_767} Y. Tanaka, T. Inami, T. Nakamura, H. Yamauchi, H. Onodera, K. Ohoyama and Y. Yamaguchi: J. Phys.: Condens. Matter {\bf 11} (1999) L505.
\bibitem{Effantin_232} J. M. Effantin, J. Rossat-Mignod, P. Burlet, H. Bartholin, S. Kunii and T. Kasuya: J. Magn. Magn. Mater. {\bf 47\&48} (1985) 145.
\bibitem{Wachter_216} P. Wachter: {\it Handbook of the Physics and Chemistry of Rare Earths}, ed. K. A. Gschneider, Jr., L. Eyring, G. H. Lander and G. R. Choppin (Elsevier Science 1994) p. 177, and references therein.
\bibitem{Clementyev_735} E. Clementyev, R. Koehler, M. Braden, J.-M. Mignot, C. Vettier, T. Matsumura and T. Suzuki: Physica B {\bf 230} (1997) 735.
\bibitem{Ott_220} H. R. Ott, B. L\"{u}thi and P. S. Wang: {\it Valence Instabilities and Related Narrow-Band Phenomena}, ed. R. D. Parks (Plenum Press 1977) p. 289.
\bibitem{Ott_772} H. R. Ott and B. L\"{u}thi: Z. Phys. B {\bf 28} (1977) 141.
\bibitem{Matsumura_217} T. Matsumura, Y. Haga, Y. Nemoto, S. Nakamura, T. Goto and T. Suzuki: Physica B {\bf 206\&207} (1995) 380.
\bibitem{Levy_364} P. M. Levy, P. Morin and P. Schmitt: Phys. Rev. Lett. {\bf 42} (1979) 1417.
\bibitem{Bucher_759} E. Bucher, K. Andres, F. J. di Salvo, J. P. Maita, A. C. Gossard, A. S. Cooper and G. W. Hull, Jr.: Phys. Rev. B {\bf 11} (1975) 500.
\bibitem{Sakakibara_790} T. Sakakibara, S. Honma, T. Tayama, K. Tenya, H. Amitsuka, T. Matsumura and T. Suzuki: Physica B {\bf 281} (2000) 566.
\bibitem{Lassailly_221} Y. Lassailly, C. Vettier, F. Holtzberg, A. Benoit and J. Flouquet: Solid State Commun. {\bf 52} (1984) 717.
\bibitem{Link_252} P. Link, A. Gukasov, J.-M. Mignot, T. Matsumura and T. Suzuki: Phys. Rev. Lett. {\bf 80} (1998) 4779.
\bibitem{Link_358} P. Link, A. Gukasov, J.-M. Mignot, T. Matsumura and T. Suzuki: Physica B {\bf 259-261} (1999) 319.
\bibitem{Link_383} P. Link, T. Matsumura, A. Gukasov, J.-M. Mignot and T. Suzuki: Physica B {\bf 281\&282} (2000) 569.
\bibitem{Mignot_424} J.-M. Mignot, P. Link, A. Gukasov, I. N. Goncharenko, T. Matsumura and T. Suzuki: Physica B {\bf 281\&282} (2000) 470.
\bibitem{Shiina_234} R. Shiina, H. Shiba and P. Thalmeier: J. Phys. Soc. Jpn. {\bf 66} (1997) 1741.
\bibitem{Shiina_779} R. Shiina, H. Shiba and O. Sakai: J. Phys. Soc. Jpn. {\bf 68} (1999) 2105.
\bibitem{Shiina_788} R. Shiina, H. Shiba and O. Sakai: J. Phys. Soc. Jpn. {\bf 68} (1999) 2390.
\bibitem{Gehring_791} G. A. Gehring and K. A. Gehring: Rep. Prog. Phys. {\bf 38} \mbox{(1975) 1.}
\bibitem{Nikolaev_793} A. V. Nikolaev and K. H. Michel: Phys. Rev. B {\bf 63} (2001) 104105/1.
\bibitem{Sakai_000} O. Sakai, private communication.

\end{thebibliography}
\end{document}